\documentclass[12pt]{article}

\usepackage{epsfig,epstopdf,latexsym,amsfonts,amsmath,amsthm,amssymb,amsbsy,multirow,slashed,wasysym,textcomp,wrapfig,graphicx,psfrag,booktabs,bbm,comment}

\usepackage[toc]{appendix}

\usepackage{color}
\usepackage{datetime}
\usepackage[
      colorlinks=false,
      linkcolor=darkblue,  
      urlcolor=blue,    
      filecolor=blue,     
      citecolor=red,
linktocpage=true,
      pdfstartview=FitV,
      bookmarksopen=true    
      ]{hyperref}

\ifpdf
\fi

\numberwithin{equation}{section}

\def\IR{\mathbb{R}}

\def\cA{{\cal A}}

\def\cO{{\cal O}}

\def\Neql#1{{\cal N}\!=\!{#1}}

\definecolor{cardinal}{rgb}{0.6,0,0}
\definecolor{darkgreen}{rgb}{0,0.5,0}
\definecolor{golden}{rgb}{0.92, 0.7, 0}
\definecolor{midnight}{rgb}{0, 0, 0.5}
\definecolor{darkblue}{rgb}{0.2, 0, 0.8}

\begin{document}

\begin{titlepage}

 \begin{flushright}
IPhT-T13/258
 \end{flushright}

\bigskip
\bigskip
\bigskip

\centerline{\Large \bf Resolving the Structure of Black Holes:}
\smallskip
\centerline{\Large \it Philosophizing with a Hammer}

\medskip
\bigskip
\bigskip
\centerline{{\bf Iosif Bena$^1$  and Nicholas P. Warner$^{1,2,3}$}}
\centerline{{\bf }}
\bigskip
\centerline{$^1$ Institut de Physique Th\'eorique, }
\centerline{CEA Saclay, CNRS-URA 2306, 91191 Gif sur Yvette, France}
\bigskip
\centerline{$^2$ Department of Physics and Astronomy}
\centerline{University of Southern California} \centerline{Los
Angeles, CA 90089, USA}
\bigskip
\centerline{$^3$ Institut des Hautes Etudes Scientifiques}
\centerline{91440 Bures-sur-Yvette, France}
\bigskip
\bigskip
\centerline{{\rm iosif.bena@cea.fr, ~warner@usc.edu} }
\bigskip
\bigskip

\begin{abstract}
\noindent We give a broad conceptual review of what we have learned about black holes and their microstate structure from the study of microstate geometries and their string theory limits.  We draw upon general relativity, supergravity, string theory and holographic field theory to extract universal ideas and structural features that we expect to be important in resolving the information problem and understanding the microstate structure of Schwarzschild and Kerr black holes.  In particular, we emphasize two conceptually and physically distinct ideas, with different underlying energy scales: a) the transition that supports the microstate structure and prevents the formation of a horizon and b) the representation of the detailed microstate structure itself in  terms of fluctuations around the transitioned state.   We also show that the supergravity mechanism that supports microstate geometries becomes, in the string theory limit,  either brane polarization or the excitation of non-Abelian degrees of freedom.  We thus argue that if any mechanism for supporting structure at the horizon scale is to be given substance within string theory then it must be some manifestation of microstate geometries.

\end{abstract}

\end{titlepage}

\tableofcontents

\section{Introduction}

The lack of black-hole hair in four-dimensional general relativity implies that  Hawking Radiation is universal and washed of all the details of the states of the matter that made the black hole in the first place.  Hawking showed that this could  lead to information loss \cite{Hawking:1976ra,Hawking:1982dj} and thereby violates some of the foundational  principles of quantum mechanics.  Over the years  there has been a vast amount of work  that either tries to address this issue head-on or tries to define the sorts of ideas or mechanisms that will be needed to resolve this paradox.

Probably the most interesting and challenging recent contribution to this discussion has  been Mathur's tightening \cite{Mathur:2009hf,Mathur:2012np,Mathur:2012dxa} of Hawking's result to show that information can only be recovered if there are either $\cO(1)$ corrections to  the Hawking states at the horizon  or  there are some other $\cO(1)$ corrections to the semi-classical physics outside black holes. This strongly suggests that information recovery from a black hole would require some macroscopic changes at the horizon and this, in turn, has stimulated a deeper discussion  of horizon-scale physics and led to some new proposals, like  {\it firewalls} \cite{Almheiri:2012rt}  (see \cite{Braunstein:2009my} and \cite{Mathur:2012jk,Susskind:2012rm,Bena:2012zi,Susskind:2012uw,Avery:2012tf,Avery:2013exa,Almheiri:2013hfa,Verlinde:2013uja,Maldacena:2013xja,Mathur:2013gua} for some  earlier and subsequent discussions).

Much of this new discussion has taken place in the framework of Quantum Mechanics and General Relativity (possibly with a cosmological constant so as to put the black hole in an $AdS$ space-time) and while there have been some interesting discussion, re-examination and recycling of old ideas, we feel that, in the absence of a mechanism to support things at  the horizon scale\footnote{And sometimes the absence of even the perception of the need for a mechanism.}, such discussions can oftentimes end up generating more heat than light. For example, at its most basic level, the arguments of \cite{Mathur:2009hf,Mathur:2012np,Mathur:2012dxa} and the firewall argument indicate that information recovery requires a black hole to have a lot of hair at the horizon.  Given the infinite blueshift at a horizon, an infalling observer would thus encounter quanta of arbitrarily high energies and would therefore be immolated.  But the no-hair theorems say there is no hair: the firewall would either fall into the black hole or be expelled from it .... and thus the stone of Sisyphus rolls back to the bottom of the hill.

In the intervening three decades since Hawking's work, we have learnt a great deal about string theory and particularly about dualities and the holographic description of field theory.  It is our view that any sensible discussion of the black-hole information problem needs to be informed by the incredible progress and body of ideas emerging from string theory.    Indeed, one of the triumphs of string theory was the work of Sen, Strominger and Vafa \cite{Sen:1995in, Strominger:1996sh} in which one could begin to see how a black hole might store the information about states and thus one might obtain  a microscopic description of black-hole entropy.  This was done for a very special class of black holes and in the limit of vanishing string or gravitational coupling.  In the 1990's this was still extraordinary because it represented real progress based upon well-defined calculations within a framework (string theory) that could provide  a viable theory of quantum gravity. 

Since the 1990's the stringy description of black holes has evolved greatly and, at the same time, open-closed string duality has given us holography and,  through it, new tools to describe strongly coupled quantum field theory.  It is thus natural that all this progress in string theory should lead to deeper developments in our understanding of black-hole microstate structure.    

A very important  body of ideas that emerged from this has become known as the {\it fuzzball program}, and while this approach is  sometimes misunderstood and mischaracterized, it is simply an attempt to understand, systematically, the microstate structure of black holes by performing calculations within string theory and, most particularly, in its low energy limit:  supergravity.  One of the very early motivations of the fuzzball program was to seek smooth, horizonless geometries that could replace not just the singularities but also the horizons of black holes with structures in which unitary evolution would be manifest.  This grew to be one of the central ideas of this program and so the results of Mathur, and the more recent firewall arguments, showing the need for new horizon-scale physics provided very gratifying confirmation of this precept.

There are roughly four different, but overlapping ways to think of fuzzballs:  
\begin{enumerate}
\item[(i)]
Follow one particular black-hole microstate as one increases the string coupling from the ``Strominger-Vafa'' regime (where the string coupling is essentially zero and the counting of states reproduces the black-hole entropy), to the regime where the supergravity description of the classical black hole is valid. One can then start seeing the effects of the back-reaction of some of the branes (while treating other types of branes as probes) and argue that in the regime of parameters where the classical black-hole solution is valid, all the Strominger-Vafa microstates become horizonless configurations.

\item[(ii)]
Use holography to analyze asymptotically-AdS black holes and study the bulk configurations that are the holographic duals of pure states of the dual CFT. Argue that none of these configurations have a horizon.

\item[(iii)] 
Use information theory arguments, as in  \cite{Mathur:2009hf,Braunstein:2009my,Almheiri:2012rt}, to argue that if information is ever to come out of a black hole and physics is to be unitary then there must be something on the right-hand side of Einstein's equations that makes a large contribution at the horizon scale. 

\item[(iv)]
Use the fact that the low-energy limit of string theory is a higher-dimensional supergravity theory with a rich massless spectrum, and construct all the possible solutions of this theory that have the same mass and charges as the black hole but have no horizon. One can then try to argue that these solutions, which go well beyond simple black holes and their apparent lack of hair, have enough entropy to account for that of the black hole.

\end{enumerate}

In Approach (i) one can imagine starting at small string coupling and studying certain classes of D-branes as one increases $g_s$. One can study this physics using the quiver quantum mechanics pioneered in \cite{Denef:2000nb,Denef:2002ru,Bates:2003vx,Denef:2007vg} and further developed in \cite{deBoer:2008zn,deBoer:2009un, Anninos:2011vn,Anninos:2012gk,Bena:2012hf,Lee:2012sc,Lee:2012naa,Manschot:2012rx,Manschot:2011xc,Manschot:2013dua,Manschot:2013sya}, and argue that the black-hole entropy comes from configurations in which the original branes that formed the black hole split into a multi-center configuration, and that certain states of this configuration grow into horizonless microstate geometries as one approaches the regime of parameters where the black-hole solution is valid. In a similar vein, one can work in an intermediate regime of parameters, where some of the branes that form the black hole have back-reacted gravitationally and some have not, and treat the latter as probes in the background sourced by the former. This ``black hole deconstruction'' uses methods pioneered in \cite{Gaiotto:2004ij,Gaiotto:2004pc} to reproduce the black-hole entropy from configurations that do not shrink and appear to have no intention of developing a horizon as $g_s$ increases \cite{Denef:2007yt,Gimon:2007mha}.

Alternatively, one can start from a D1-D5-P black-hole microstate at weak coupling and compute the string theory amplitude that corresponds to emitting a closed string \cite{Giusto:2011fy}. This amplitude is non-zero in certain channels, and these channels, which differ from microstate to microstate, give the extra dipole moments of the back-reacted microstates. Since these extra dipole moments are all zero for the classical black-hole solution, this indicates that the back-reacted Strominger-Vafa microstates do not have a horizon. Knowing the kinds of dipole moments that typical states will source is also a useful hint in searching for their exact back-reacted solution.

Approach (ii) clearly works in holographic systems that do not have a large classical black hole in the bulk, such as in the LLM solutions and their M2 brane limits \cite{Lin:2004nb,Bena:2004jw}, the D1-D5 system \cite{Lunin:2001fv, Lunin:2002iz, Kanitscheider:2007wq,Skenderis:2008qn}, or Polchinski-Strassler (PS) backgrounds \cite{Polchinski:2000uf}. In all these examples there exists a very symmetric, singular solution that can be thought of as giving the  ``thermodynamic'' description of the physics. However, none of the pure states of the boundary theory is dual to this singular solution, they are rather dual to smooth solutions that differ from the naive, symmetric one at scales parametrically larger than the Planck size. If one extrapolates how holography works in these systems to holographic configurations that have a large black hole in the bulk, one can argue that there is nothing holy about singularities behind horizons and that the classical black-hole solution is the analogue of the singular LLM, D1-D5 or PS geometries, and therefore should not be dual to any pure state. Moreover, the configurations dual to pure states are expected to differ from the naive solution at scales parametrically larger than the Planck size, and they are therefore naturally expected to be horizonless.

Another way of realizing the second approach is to argue in the context of the AdS-CFT correspondence that the boundary theory cannot reconstruct the spacetime inside the black hole, or that, even if it is reconstructed, the quantum fluctuations around it will be so large that the notion of spacetime will stop making sense. Arguments both for and against this interpretation have appeared in \cite{Kraus:2002iv,Balasubramanian:2005mg,Papadodimas:2012aq,Almheiri:2013hfa,Papadodimas:2013wnh,Papadodimas:2013jku,Verlinde:2013qya}.

Approach (iii) has many names, it has been called fuzzballs, firewalls, energetic curtains, and consists of using information-theory arguments to demonstrate that the black-hole horizon should be replaced by something else if one wants information to come out of a black hole and physics to be unitary. 

Both approaches (ii) and (iii) can be thought of as ``bottom-up'' approaches in that they argue that something must replace the black hole and emerge at the horizon scale, but say very little about what  that something is. It is not clear how much of the problem with this inability is intrinsically built in, and how much can be overcome with enough work and insight. For example, one can imagine trying to use holography to disentangle between two bulk solutions dual to two different typical microstates of the D1-D5-P black hole, and argue that since the vev's in these microstates differ by corrections of order $1/{N_1 N_5}$, these bulk microstates should both have black-hole-like throats that cap off at a length of order $N_1 N_5$ to some power, and the different structures of the cap are therefore the cause of the differences in the vev's. However if one includes quantum effects in the calculation of these vev's one also obtains effects of this order, and hence it is hard to figure out whether the difference of the two microstates comes from quantum fluctuations on top of the black-hole background or from two bulk solutions with long throats and different caps \cite{Skenderis:2007yb}.
  
Approach (iv) takes the opposite, or ``top-down,'' route by starting with string theory, or supergravity, and constructing horizonless solutions that have the same mass, charge, angular momentum and size as a black hole, and exist in the same regime of parameters where the classical black hole has a macroscopically-large horizon area. An (incomplete) list of solutions that have the same types of charges as these black holes can be found in \cite{Mathur:2003hj,Giusto:2004id, Bena:2005va, Berglund:2005vb, Saxena:2005uk,Ford:2006yb, Bena:2007kg, Bena:2010gg,Giusto:2011fy, Lunin:2012gp, Giusto:2012yz}.  

Since quantizing strings in curved backgrounds that have Ramond-Ramond fields with about a dozen different orientations is, at present, out of reach of current technology, practical considerations mean that most of the configurations that have been constructed are smooth horizonless supergravity solutions.  These have come to be known as {\it microstate geometries}. \cite{Bena:2004de,Bena:2005va, Berglund:2005vb,Bena:2007kg}. Working in higher-dimensional supergravity theories already allows one to see what new features emerge and how one can evade the singularity theorems and ``no go'' theorems of four-dimensional General Relativity. Furthermore, by focusing on smooth horizonless solutions whose curvature is much larger than the Planck scale, one ensures that $g_s, \alpha'$ and quantum corrections will not wipe out the solutions (although, as we will see in Section \ref{Sect:Quantum}, things are not always as simple as this).

However, working purely within supergravity is not enough: for example supergravity does not have a way to make sense out of certain singularities that are benign in string theory, and moreover one can use string dualities to transform a smooth supergravity solution into a singular one. Both solutions make perfect sense in string theory, but if one is only willing to accept smooth supergravity solutions one is forced to throw away one solution and keep its dual;  to us this appears very hard to justify on physical grounds (although some attempts have been made in \cite{Sen:2009bm}). Moreover, there exist multi-center solutions for which each center gives a smooth patch of spacetime in a certain duality frame, but there is no frame where the metric near all the centers is smooth \cite{Vasilakis:2011ki}. 

In addition, when making smooth supergravity solutions using fluxes threading cycles, one has the choice of how many flux quanta to put on a given cycle. When there are very few such quanta, the size of this corresponding cycle may become very small, and can thus approach the Planck scale. In this limit, such a solution cannot be accurately described by the supergravity approximation but it still represents a state of the system and so cannot simply be discarded.

Given these considerations, we must enlarge the scope of our discussion  beyond microstate geometries so as to encompass both the singular solutions described above and the Planck-scale configurations that lie at, or beyond, some well-defined physical limit of supergravity.  We will continue to reserve the term {\it microstate geometries}  to describe the smooth horizonless solutions that are described fully within supergravity and we will introduce a broader category of  {\it microstate solutions} to connote  physically meaningful string/D-brane backgrounds in the sense outlined above.  It is hard to give a rigorous definition of this  notion of {\it microstate solutions}   but, as we will discuss in Section \ref{Sect:Mechanism}, there are two (related) ingredients that will most certainly enter in this definition. The first is that near each place where the solution is singular, the singularity should look like that of a D-brane, or another 16-supercharge object. The second is that there should exist a duality sequence that takes each patch of the solution that has a singularity into a smooth patch. 

Hence, we make the following definitions:

\begin{enumerate}

\item[1.] A {\em microstate geometry} is a smooth horizonless solution of supergravity that is valid within the supergravity approximation to string theory and that has the same mass, charge and angular momentum as a given black hole.

\item[2.] A {\em microstate solution} is a horizonless solution of supergravity, or a horizonless, physical limit of a supergravity solution,  that has the same mass, charge and angular momentum as a given black hole.  Microstate solutions are allowed to have singularities that either correspond to D-brane sources or can be patch-wise dualized into a smooth solution.  

\item[3.] A {\em fuzzball} is the most generic horizonless configuration in string theory that has the same mass, charge and angular momentum as a given black hole. It can be arbitrarily quantum and arbitrarily strongly curved. 

\end{enumerate}

Hence, one can think of microstate geometries as fuzzballs that can be described as smooth, horizonless geometries within the validity of the supergravity approximation and one can think of microstate solutions as being valid string theory backgrounds that can be constructed and analyzed using supergravity and (current) string theory technology.  While our particular efforts have largely focussed upon microstate geometries, our purpose here is to take what we have learned from the study and construction of microstate geometries and  try to extract some of the universal ideas that we believe important to resolving the microstate structure of any black hole, including Kerr and Schwarzschild.   This exercise will thus take us into the more speculative arenas of microstate solutions and generic fuzzballs.

Section 2  of this paper contains some of the important ideas that have emerged from string theory that have changed the way in which we must think about black holes.   Section 3 shows that  microstate geometries use the only possible supergravity mechanism for supporting time-independent horizonless solitons and highlights the crucial role of geometric transitions and non-trivial  topology in the space-time.    Further insights from holographic field theories are outlined in Section 4 and most particularly it is argued that a proper understanding of black-hole physics involves the emergence of a transition scale and the realization of an energy gap.  In Section 5 we discuss various string theory limits of microstate geometries and how they are related to brane polarization or the appearance of non-Abelian degrees of freedom.  In Section 6 we consider methods by which one can transfer results and ideas from the BPS story to near-BPS and near-extremal systems.  Section 7 makes a brief foray into some of the  quantum properties of microstate geometries and discusses why the transition scale of such geometries might well be large.  Finally, in Sections 8 and 9, we try to draw out some  of the broader conclusions about the microstate structure of Schwarzschild or Kerr black holes.

Many of these important ideas arise, sometimes in retrospect, from the manner in which fuzzballs and microstate geometries evade a remarkable collection of ``no go'' theorems.  The result amply illustrates Nietzsche's aphorism:  ``What does not destroy me, makes me stronger.''  In the same vein, our general approach in this paper will be motivated by the subtitle from the source of this quote:  {\it How to Philosophize with a Hammer}\footnote{While we cannot pretend to the brilliance of Nietzsche, our purpose  is similar to his in writing {\it Twilight of the Idols}:  to provide a broad introduction to our thoughts and work.  It is also amusing to note that, like Nietzsche's book, the first draft of this paper was written in exactly the same week, precisely 125 years later.}.

\section{An Early Lesson from String Theory: \hfill \break The Horowitz-Polchinski correspondence point}

From the perspective of string theory, microstate solutions appear to be impossible. Indeed, while the microstates of a black hole had been successfully counted at vanishing string coupling \cite{Strominger:1996sh} it seemed certain that they would have to disappear inside the horizon at any finite value of the string coupling \cite{Horowitz:1996nw,Damour:1999aw}.  Perturbative string states, like all normal matter, will become more and more compressed as the string, and hence gravitational, coupling increases.  On the other hand, the size of a black-hole event horizon increases with the gravitational coupling and so there is a cross-over (the Horowitz-Polchinski correspondence point), at very small string coupling, where the stringy microstate structure gets swallowed by the horizon.  

However, string theory comes equipped with some classically abnormal matter:  D-branes and NS5-branes.  The important point is that D-branes are solitonic objects and their tension is proportional to $g_s^{-1}$, while the tension of NS5-branes is proportional to $g_s^{-2}$.  This means that they become floppier and floppier as the string coupling increases and so momentum modes, open string modes and other branes stretched between or dissolved in the right combinations of branes can give rise to configurations that grow, with $g_s$, at exactly the same rate as the would-be horizon size.  Thus microstates made out of branes can avoid sinking into the Great Grimpen Mire of horizons.

Isolated clusters of  branes typically result in singular sources for the Maxwell fields and in singular microstate solutions, and to get smooth microstate geometries one must replace the branes by smooth cohomological fluxes.  We will discuss this more in Section \ref{Sect:Supergravity} but we note here that even after such a geometric transition to cohomological fluxes, the supergravity solution will have the same growth with $g_s$ as the black-hole horizon, and thus evade the Horowitz-Polchinski crossover with a horizon.

There are further consequences of this proper growth with $g_s$ that can be used to sharpen our ideas of what constitute microstate solutions.  One of the common misconceptions about microstate solutions is that they really only exist for BPS configurations and involve some kind of force balance between gravity and electrostatic repulsion. This simple picture is correct if one tries to break a black hole into all the branes that make it, and spread these (singular) brane sources on the Coulomb branch. However, the smooth microstate geometries we discuss have no singular sources -- all their charge comes from  cohomological fluxes -- and the forces involved in their stabilization are primarily magnetic. This significantly modifies the intuitive picture and perturbative stability of BPS configurations. One way to see this is that as one varies $g_s$ or the Newton constant, the size of these geometries changes, and this does not happen for Coulomb branch solutions. Hence, as we will discuss in more detail in Sections \ref{Sect:Misconceptions} and  \ref{Sect:NearBPS}, we do not expect the smooth microstate geometries to be destroyed by non-extremality.  The major reason why, at this point, most of the microstate geometries that have been constructed are BPS is technical:  supersymmetry makes the equations linear and relatively easier to solve.

\section{Lessons from Supergravity}
\label{Sect:Supergravity}

The supergravity theories that are usually used in the construction of black holes in asymptotically flat space-times are the massless, ungauged supergravity theories that come from the dimensional reduction of M-theory or type IIB and IIA supergravity on compact Ricci-flat  manifolds.  The bosonic sector of such theories consists of the graviton coupled to scalar fields and tensor gauge fields.  We will therefore focus on such theories here but our conclusions will certainly admit interesting generalizations in other contexts.

\subsection{Topological stars}

There are many theorems that prove that there are no solitons in these ungauged, massless supergravity theories.  The intuition is rather simple:  Massless fields travel at the speed of light and the only way to hold such fields into a classical lump is to have a black hole and a horizon.  This leads to the conventional wisdom expressed in the mantra that there are ``No Solitons without Horizons''  (see, for example, \cite{Gauntlett:1998fz,Shiromizu:2012hb}). It therefore might seem mysterious that there can be  microstate geometries since these theorems would appear to forbid them but, once again, this is does not happen  \cite{Gibbons:2013tqa}.

\subsubsection{Microstate geometries and some common misconceptions}
\label{Sect:Misconceptions}

A smooth microstate geometry is a horizonless solution that looks like a black hole from a distance and does not have an interior:  the space-time caps off before the would-be horizon forms.  

There are several misconceptions that frequently appear in discussions about these objects.  First there is the belief that these solutions involve matter that somehow floats above a horizon.  Second, there is sometimes the assumption that there is still a remnant of the interior of the black hole. Third,  is the belief that, in the collapse, an event horizon is formed and then the transition to a fuzzball state involves some kind of process inside the star that involves replacing the horizon with fuzz.  

First and foremost, one of the central points of fuzzballs is that there is no horizon and there never was a horizon. The presence of a horizon implies that there is entropy whose microstate structure we cannot see  and, as Hawking and Mathur have so amply demonstrated, horizons result in information loss.   So microstate geometries do not have horizons, never had horizons, and (hopefully) can never evolve into things that have horizons.  Thus there is no need of a mechanism to enable matter to float above a horizon.  There is, however, the need for a phase transition to a state of massless matter that is ``stiff enough'' to prevent collapse into a black hole.  As we will see, magnetic fluxes and Chern-Simons terms are indeed strong enough to hold up such an object against collapse into a black hole   and this suggests  the possibility of  a new class of objects in Nature:  {\it Topological Stars}.  In the fuzzball program we would like to think of these as simply being a more precise description of black holes.

Since there is no horizon there is no obvious analogue of the ``inside'' of the horizon in a microstate geometry.  It may therefore seem rather difficult to imagine how such an object could behave like a black hole and trap infalling matter and radiation. However, microstate geometries do have  an immensely complicated configuration of branes and fluxes that has replaced the matter that made the original star.  For the D1-D5 system, where all the microstate solutions are known but there is no black hole, one can see how incoming particles that scatter in a certain horizonless solution can end up spending a lot of time inside this solution before getting out, in agreement with the dual CFT \cite{Lunin:2001jy}. 

Going further, one can argue on generic grounds that microstate geometries provide a particular (coherent state) basis of the Hilbert space of a certain black hole, but that a typical state of this system will be in general some random vector in the Hilbert space that does not have a geometric meaning because it  is a quantum superpositions of many geometries. To compute the scattering of a given particle off such a state one has to compute the scattering off each of the geometric microstates that span it, and add the corresponding amplitudes. Since all microstates are different, they will not scatter this particle in the same direction, and hence adding the resulting amplitudes will give a very small number. For a random vector in a Hilbert space that is spanned by $e^S$ microstate geometries, the resulting amplitude will be given by the length of a random walk with $e^S$ states, which is $e^{S/2}$, and hence the incoming particle will be absorbed and not re-emitted in its original form before the black hole evaporates completely.  Hence, infalling particles become trapped inside the microstate structure because of (Anderson) localization in a region with a very large number of states. The information in these particle will affect the structure of these states and consequently will change the details of the Hawking radiation emitted from the typical state.  In this way a microstate geometry can replicate the macroscopic behaviour  of a black hole but without  information loss.  From this perspective, a topological star is little different from a neutron star:  one is simply scattering off an exotic state of matter. 

On the other hand, it has been argued that because of the density of states in a black hole and in a microstate geometry, certain classes of observers will make observations of the infall that are consistent with the classical experience of falling across a horizon and hitting the singularity.  We stress that this experience is simply a way in which some observers might perceive  interactions with fuzzball states.  This possibility underpins the idea of {\it fuzzball complementarity} \cite{Mathur:2010kx,Mathur:2011wg,Mathur:2012jk,Mathur:2012zp} which necessarily relies upon strong quantum effects within the very high density of states within the black hole. We will comment further on this in Section \ref{Sect:Fire} but since it is hard to upgrade these heuristic arguments to more precise computations we  remain agnostic on this very interesting suggestion.

The idea that an event horizon forms during the collapse of a shell, to be subsequently wiped out by a fuzzball is sometimes used as a straw-man to argue against fuzzballs. Indeed, if the fuzzball degrees of freedom are not kicking in just before  a horizon would form, all degrees of freedom excited during the subsequent evolution of the shell will not be able to affect the horizon because they will be in its future, and therefore cannot change the physics backwards in time. Thus one  expects, based upon classical physics, that the fuzzball transition occurs when the collapsing matter approaches the would-be horizon scale. 

However, as we will discuss in section \ref{Sect:Quantum}, there are some subtleties. In particular, as was argued in \cite{Mathur:2008kg}, we expect to have $e^S$ fuzzball states into which the shell can tunnel, and even if the probably of tunneling into one particular one is small, of order $e^{-S}$, one can still estimate that the shell will tunnel with probability one into a fuzzball. Thus the transition to the fuzzball state will generically involve a quantum transition, particularly if the initial conditions of the infall are tuned to have a high level of symmetry.  Moreover, as we will explain in Section \ref{Sect:Mechanism}, the reason why fuzzballs can have a large size is because they have low-mass degrees of freedom, which affect the physics on horizon scales. By the uncertainty principle, we expect the quantum effects corresponding to these low-mass degrees of freedom to have a time uncertainty of order the horizon crossing time, and this quantum uncertainty might as well affect the physics backwards in time from the singularity.  To the extent that the physics of microstate solutions suggests that quantum mechanics is becoming important on the horizon scale, the backwards-in-time resolutions  of singularities on  scales of order the infall time might not be such an outrageous proposition.

\subsubsection{No Solitons without topology}
\label{Sect:topology}

The simplest  ``no-go'' theorems that purport to deny the existence of horizonless solitons (see for example \cite{Gauntlett:1998fz,Shiromizu:2012hb}) start merely from the assumption that the space-time is stationary and asymptotically flat and that the matter fields are time-independent.   One can then use the time-like Killing vector in the Komar formula to write an expression for the ADM mass and one can then convert this to an integral over space-like hypersurfaces.  If one a) assumes that these surfaces have no singularities, b)  neglects Chern-Simons interactions and c) neglects all topology on the hypersurfaces, then the Einstein equations enable one to recast this spatial volume integral as a total divergence and conclude that $M=0$.  This means that the space-time must be globally flat.  
 
However the standard theorems had either neglected the Chern-Simons interactions that typically appear in supergravity theories or had assumed that there was no non-trivial topology on space-like hypersurfaces.  It was shown in  \cite{Gibbons:2013tqa}\footnote{See \cite{Kunduri:2013vka} for extensions of this idea.} that both Chern-Simons terms and topology can potentially contribute to the ADM mass  and thus allow for the existence of solitons.  This was carefully analyzed for five-dimensional $\Neql 2$ supergravity coupled to vector multiplets and while the possible Chern-Simons contributions actually cancel, there was still a topological contribution to the mass. Indeed, let $K$ denote the Killing vector, $F^I$ denote the Maxwell fields and define $G_I$ to be the $3$-form duals of the $F^I$:
\begin{equation}
G_I  ~\equiv~  *_5 (Q_{IJ} \, F^{J}) \,,     \label{Gdefns} 
\end{equation}
where $Q_{IJ}$ is the matrix of scalar fields in the kinetic terms of the Maxwell fields. The interior product, $i_K(G_I)$, is a two-form and if there are non-trivial $2$-cycles then $i_K(G_I)$ could have a non-trivial harmonic part, $H_I$.  One can then show that the ADM mass is given by an expression of the form:
\begin{equation}
M  ~\sim~     \int_{ \Sigma}   \, H_{I}  \wedge F^I  \label{Mtop} \,,
\end{equation}
where $\Sigma$ is a global Cauchy surface. Note that since $H_I$ is harmonic, this integral also extracts the harmonic part of $F^I$ and so the integral is entirely topological, reducing to the intersection form on the cohomology.  

The Chern-Simons terms gives rise to the electric charge of the solution:
\begin{equation}
Q_I  ~\sim~   C_{IJK}\,  \int_{ \Sigma}   \, F^J  \wedge F^K  \label{Qtop} \,.
\end{equation}

There are several important conclusions to be drawn from this result.  

First, one must remember that the only assumption, apart from from the absence of horizons, smooth geometry and asymptotic flatness, was that the space-time and matter should be time-independent.  Therefore, the {\it only} way to support a stationary microstate geometry, even a Schwarzschild one, in these supergravity theories is with cohomological fluxes. This means that the ``force balance'' underlying microstate geometries is not the usual Majumdar-Papapetrou type of balance between electric charge repulsion and gravitational attraction that one encounters in Coulomb-branch solutions. Instead, it is a balance of {\it magnetic} fluxes that tend to expand a topological cycle and gravity that tends to collapse this cycle.  The electric charges of the black hole only emerge from the interactions between these magnetic fluxes.

For BPS solutions the  spatial projections of the cohomological fluxes are all either self-dual or anti-self-dual and one can show directly that these formulae yield precisely $M \ge \sum_I | Q_I | $.  For generic non-BPS solutions one can decompose the spatial projections of the cohomological  Maxwell fields into self-dual and anti-self-dual parts, then the electrostatic charges are essentially given by the difference of the squared magnitudes  of the self-dual fluxes and the anti-self-dual fluxes.  If one makes some assumptions about the form $H_I$ then one can show that the mass is given by the {\it sum} of these two contributions and thus argue that,  in general, one has $M \ge \sum_I | Q_I | $.

The important conclusion here is that besides Chern-Simons terms and cohomological fluxes, there is no other mechanism  that can support a stationary microstate geometry. In a generic black hole this will require a mixture of self-dual and anti-self dual fluxes and so if one is to construct a stationary Schwarzschild microstate this would require new classes of long-lived ``flux-anti-flux'' solutions, similar to those conjectured in \cite{Aganagic:2006ex} for flux compactifications.

This could  present a very interesting set of new challenges to the solution generating techniques that come from inverse scattering methods.  Based upon the structure of BPS solutions, we expect to find very large families of axisymmetric, non-extremal microstate geometries, which depend on several functions of two variables. Their construction can be thus reduced to an effectively two-dimensional problem that should be amenable to inverse scattering methods applied to the corresponding equivalent scalar coset models \cite{Figueras:2009mc, Katsimpouri:2012ky}. However, as we have seen, the construction of these geometries critically depends upon the existence of  topological terms, whose incorporation in this technology tends to be rather difficult to do systematically. Once this barrier is overcome, we expect finding these rich families of axisymmetric, non-extremal microstate geometries to be within reach. 

Finally, independent of the entire fuzzball proposal, this supergravity analysis suggests another possible  end-point of stellar collapse: {\it Topological Stars}, supported by higher-dimensional magnetic fluxes.  It is important to realize that the topological cycles necessarily require extra, compactified dimensions of space-time and so the structure of such a star can only appear at the scale of the extra dimensions.  Through the effects of warp factors, it is possible that the scale of  (some of) these extra dimensions could become much larger near the star than they are at infinity.  The fact that these stars are made of flux and anti-flux suggests that they will slowly decay. In the fuzzball program we  suspect that this will be how fuzzballs generate Hawking radiation but, more generally, it will mean that  topological stars will still emit radiation.  We will discuss this further in Section \ref{SKmicrostates}.

\subsection{Geometric transitions and new black-hole scales}
\label{GT1}

In order to characterize the microstate solutions, and to understand the features one expects from them if they are to describe the typical microstates of the black hole, it is useful to discuss several scales one must encounter in their physics. First, one can define the scale, $\lambda_T$, of the geometric transition that is responsible for replacing certain singular brane sources with smooth geometry.  We expect this scale to be the typical size of the homology cycles, or bubbles in a microstate solution and it will be  determined by a balance between gravity that tends to collapse the bubbles and the flux that tends to drive their expansion.  For BPS and extremal bubbles we typically find that 
\begin{equation}
\lambda_T ~\sim~ k \, \ell_P 
\end{equation}
 where  $k$ is the number of flux quanta threading the typical cycle and  $\ell_P$ is the Planck length\footnote{For details, see, for example,  \cite{Bena:2007kg}.  In a BPS microstate geometry with a long throat that corresponds to a four-dimensional inter-center distance of order $d$, the warp factors scale as  $Z \sim k^2/d, V \sim1/d$ and hence $\lambda_T^2 \sim d^2 (Z_1 Z_2 Z_3)^{1/3} V$.}. 
 
 After the geometric transition, the entire region where the black hole would have formed will be populated by a cluster of homology cycles and fluxes whose overall size is approximately that of what would have been the black hole. Thus one would expect that the number of cycles,  $N$, in the transitioned geometry will be approximately the number of bubbles that can be stacked across a configuration of order the would-be horizon area. Based upon dimensional analysis, for a black hole in $D$ space-time dimensions we expect:
\begin{equation}
N ~\sim~ \Big( \frac{\cA^\frac{1}{D-2}}{\lambda_T} \Big)^{D-1} \,,
\end{equation}
where $\cA$ is the horizon ``area'' of the would-be black hole.

For BPS, extremal and near-extremal microstates one can also define another scale: the length of the throat. It is well known that BPS black holes have throats of infinite proper length, and the microstate geometries of these black holes have very long but finite throats, that end in a smooth cap \cite{Bena:2006kb,Bena:2007qc}.  For near-extremal black holes, that have long but finite throats, the length of the throat of microstate solutions can be compared to the length of the black hole throat, to determine how close the microstate solution is to describing typical states of the black hole \cite{Bena:2012zi}.  Extremal and near-extremal black hole throats can also be characterized by their width (proportional to a power of the horizon area) and the throats of microstate solutions are expected to have the same width. For extremal solutions this width is controlled only by the charges, so the matching is automatic.

The length of the throat determines another more universal and physically important quantity: the maximum redshift, $z_{max}$, between the energy of an excitation measured at the cap and the energy measured at infinity.  General, non-extremal microstate geometries do not have $AdS$ throats but we still  expect them to be ``capped off'' before a horizon can form and the region around the microstate structure will be compact and localized. One therefore expects that excitations of this structure will be gapped with a minimum energy splitting, $E_{gap}$, measured at infinity and that there will not be a continuous spectrum. 

The longest wavelength of an excitation of the microstate structure will be of order the would-be horizon scale, $\frac{2 G M}{c^2}$, and the lowest energy gap will simply be that of a wave with this wavelength redshifted by $z_{max}$.   This red-shifted wavelength will generically be much longer than $\frac{2 G M}{c^2}$, and hence have an energy much less than the typical Hawking quanta.   We therefore expect a generic (non-extremal) black-hole microstate to involve at least two physical parameters:  the scale of the non-trivial topological cycles that support it and the energy gap of the spectrum measured at infinity.   The latter will also be related by the red-shift, $z_{max}$, to the lowest-energy  collective modes of the fuzzball. For the classical black hole horizon this redshift is infinite and the spectrum is not gapped. Hence, $z_{max}$, or alternatively $E_{gap}$, indicate how much a given microstate geometry differs from the classical black hole.

Thus, besides the Planck length and the horizon size, a microstate geometry will always come with two completely new physical scales:  the scale of the geometric transition, or the size of the bubbles, $\lambda_T$, and the energy gap in the spectrum,  $E_{gap}$. The two new scales should play a crucial part in the resolution of the information paradox.  One should note that these two scales also govern rather different physics:  The transition scale provides the mechanism to support the geometry while $E_{gap}$ governs fluctuations about the transitioned geometry.  

Having established a mechanism and the role of $\lambda_T$, there are important and distinct follow-up questions as to the extent to which semi-classical geometries can capture details of the typical microstates of a certain black hole, and some of these questions boil down essentially to whether $\lambda_T$ is larger than the Planck length or not. In the absence of a detailed counting of microstate geometries that reproduces the black-hole entropy, we do not know what this scale will be for the typical microstate geometries of extremal black hole, and we know much less about  non-extremal microstate geometries. However, in Section \ref{Sect:Holography} we will examine  the interpretation of these scales in the dual holographic theory and discuss the extent to which holography can tell us anything about them. Moreover, we will discuss in Section \ref{EntFavoring} why one might expect $\lambda_T$ to be large.

Before finishing, we would like to emphasize, once again, an important physical point about microstate geometries, microstate solutions and fuzzballs.  The issue of how one supports such an object is  distinct from how one describes the typical microstates.  This is reflected in the independence of the two new scales and the physics they govern.    More generally, in Section \ref{Sect:Mechanism}, we will discuss other seemingly distinct mechanisms for supporting generic microstate solutions and  fuzzballs and argue that that they are simply different incarnations of the fundamental mechanism described here but that emerge in different regimes determined by the values of $\lambda_T$ and  $E_{gap}$.

\subsection{Examples and some other no-go theorems}

There are vast families of explicit examples of BPS microstate geometries and of the geometric transitions that generate them (see, for example, \cite{Mathur:2003hj,Giusto:2004id, Bena:2005va, Berglund:2005vb, Saxena:2005uk,Ford:2006yb, Bena:2007kg, Bena:2010gg,Giusto:2011fy, Lunin:2012gp, Giusto:2012yz}).  There are also more limited classes of extremal,  non-BPS examples and a handful of non-extremal microstate geometries \cite{Jejjala:2005yu,Bena:2009qv,Bobev:2009kn,Compere:2009iy,Dall'Agata:2010dy,Bobev:2011kk,Bossard:2011kz,Bossard:2012ge,Niehoff:2013mla}.

In five dimensions, the three-charge BPS microstate geometries  exemplify the mechanism outlined above.  Interestingly enough, there could have been  major impediments to the existence and construction of even the BPS geometries.  

First, the BPS solutions can be written as a time-fibration over a four-dimensional base whose metric is conformal to a hyper-K\"ahler metric \cite{Gauntlett:2003fk,Gutowski:2004yv,Bena:2004de}. However, the only four-dimensional Riemannian hyper-K\"ahler metric that is asymptotically flat is the globally flat metric on $\IR^4$ and so there can be no topology to support the fluxes that are essential to microstate geometries.  This  ``no-go'' theorem  is evaded by a remarkable fact first discovered in \cite{Giusto:2004kj} and then extensively generalized and exploited in \cite{Bena:2005va, Berglund:2005vb}:  The hyper-K\"ahler metric on the spatial base does not have to be Riemannian, it can be ambi-polar, which means that the signature can change from $+4$ to $-4$, and the conformal factor can be arranged so as to compensate and create a completely smooth five-dimensional, Lorentzian metric.

Another potential problem with the three-charge  microstate geometries was that the presence of three independent electromagnetic charges means that the Chern-Simons interactions will enter non-trivially into both the equations of motion and the BPS equations.  While we have shown that this is essential for the existence of microstate geometries, the non-linearity that it introduces could have made the equations impossible to solve in practice, which would have made the phase-space structure impossibly difficult to analyze.  However, another miracle happens: the BPS system in five dimensions is actually linear  \cite{Bena:2004de} and the non-linearities of the equations only appear in the sources on the right-hand sides of the equations.  The entire system could thus be reduced to four-dimensional, Euclidean electromagnetism.  This enabled incredible progress in the analysis and understanding of what had, at first sight, seemed to be an impossibly difficult problem.

Relatively recent work has shown that the corresponding BPS equations in six dimensions \cite{Gutowski:2003rg,Cariglia:2004kk} are also largely linear \cite{Bena:2011dd,Giusto:2013rxa}.  It is expected that this will lead to similar advances in the study of six dimensional BPS microstate geometries.  While this may sound like an incremental improvement, there is a conjecture that dramatic new BPS structures will exist in six dimensions \cite{Bena:2011uw} and the linearity of the BPS equations will play a crucial role in understanding these structures.
 
The fact that microstate geometries have dodged so many potentially lethal challenges to their existence suggests that they are remarkable mathematical structures and leads us to hope that they will indeed find an important place in the natural world.

\section{Lessons from Holography}
\label{Sect:Holography}


There are two basic ways in which one can use holography to study black-hole microstates.  The most universal method is to take a standard  black hole and place it in an $AdS$ background.  The AdS/CFT correspondence then implies that there should be a description of the black-hole microstates in terms of the dual CFT and that the evolution of the black hole should be described by the unitary evolution in that CFT.   The simplest incarnation of this is the duality between black holes in $AdS_5$ and strongly coupled $\Neql4$ Yang Mills theory.  This new perspective was one of the early driving forces for the growing belief in the 1990's that black-hole evolution should be unitary and that there should not be any information loss. 

BPS and extremal black holes and black rings make the connection with a dual CFT even more directly.  Such black holes have a near-horizon limit that is of the form $AdS_3 \times S^{D-3}$ or $AdS_2 \times S^{D-2}$ and so are dual to a CFT in 1+1 or 0+1 dimensions.  The original counting argument of  \cite{Strominger:1996sh} involved a black hole constructed from momentum modes on a combination of D1 and D5 branes in the IIB string theory.  This creates a black string in six dimensions whose throat is asymptotic to  $AdS_3  \times S^{3}$.  The spatial direction of the dual $(1+1)$-dimensional CFT is the common D1--D5 direction and this CFT has central charge $c = 6\, N_1 N_5$, where the $N_1, N_5$ denote the number of D1 and D5 branes.  The state counting is done by enumerating the ways in which momentum can be added to this CFT.

\subsection{The energy gap and typicality}
\label{Sect:Typicality}

When the size, $L$, of the D1-D5 common direction is finite, the CFT describing the D1-D5-P system is put in a box and has a natural gap energy, $E_{gap} \sim \frac{1}{N_1 N_5 L}$, either defined by the lowest-dimension operator in the theory or given by the smallest energy gap between states in the theory.  In the D1-D5 system this is given by the dimensions of twist operators associated with the orbifolding of identical strands of the D1-D5 system.  Alternatively, these twist operators intercommute the strands of the D1's and D5's, to yield a ``long effective string'' of maximal length $N_1 N_5 L$ and the lowest momentum mode on this string has the energy $E_{gap} \sim \frac{1}{N_1 N_5 L}$. More generally, in any conformal field theory one expects the gap energy to scale as  $E_{gap} \sim c^{-1}$, where $c$ is the central charge.  There are, of course, sectors that have shorter effective strings, but the sector with the longest such string  contains the most entropy because the gap is the smallest.  For this reason, this is usually called the ``typical sector.''

We now have two definitions of ``the gap:'' the one given here and the gravitational description in terms of red-shifts and collective excitations given in Section \ref{GT1}.  Obviously we would like these definitions to coincide and for BPS black holes one can establish that this follows from the semi-classical quantization of the moduli space of microstate geometries. 

To understand how this comes about, we first note that each bubble of geometry comes with an intrinsic angular momentum as seen in four dimensions \cite{Berglund:2005vb}.  This arises, once again, from the Chern-Simons interaction:  Two magnetic fluxes combine to make a source of electric field and that electric field combines with the third magnetic field to create an angular momentum through an $\vec E \times \vec B$ term, whose direction depends upon the spatial arrangement of the bubble.  Semi-classical quantization of the moduli space implies that all these angular momenta must be individually quantized  \cite{Bena:2006kb,Bena:2007qc,deBoer:2008zn,deBoer:2009un}.  On the other hand, at least for BPS microstate geometries, the depth of the AdS  throat is an extremely sensitive function of the moduli of the solution, and specifically, of the relative angles between the cycles. The semi-classical quantization of the angular momenta  limits the fine tuning of the bubble moduli and hence limits the depth of the black-hole-like throat.  This then yields the maximal redshift that a given microstate geometry can have, which can then be fed into the calculation in Section \ref{GT1}.  This produces the same value of $E_{gap}$ as the one arising from the typical sector of the conformal field theory.

This computation has several important consequences.  
\begin{itemize}
\item It shows that the semi-classical microstate geometries are dual to states that belong to the typical sector of the dual conformal field theory and not some strange, highly-specialized outliers.  Thus microstate geometries sample the most important thermodynamic sector of the theory.
\item Even though the computation has only be done for the BPS sector, it strongly supports the realization of $E_{gap}$ for generic black holes in terms of maximal redshifts of longest wavelengths as described in Section \ref{GT1}.
\item It shows that quantization of the moduli space and changes of order $\hbar$ can wipe out very large scale regions of space-time in which the supergravity approximation is valid.
\end{itemize}
The first of these points obviously builds confidence in the efficacy of the microstate geometries in sampling the black-hole microstates and the second gives us another feature that we believe will generalize to microstate geometries of non-extremal objects.  The third point is a remarkable phenomenon \cite{Bena:2007qc,deBoer:2008zn,deBoer:2009un} that notes that by classical tuning one can make ``abysses''  \cite{Bena:2007qc} in which the throat of a BPS microstate geometry becomes arbitrarily deep. This geometry can consist of large bubbles where the curvatures are extremely small and supergravity approximation apparently well within its range of validity ... and yet a change of order $\hbar$ can wipe out may ``cubic'' megaparsecs in such an abyss.  This means that despite their vast size and the validity of supergravity, such geometries only have a phase-space volume of order $\hbar$ and they can be wiped out by a single quantum fluctuation.  We will comment further on this in Section  \ref{Sect:Quantum}.  

\subsection{IR physics and the dynamical generation of the transition scale}

Holographic field theory is replete with examples in which singular, symmetric but unphysical solutions are replaced by less symmetric solutions that can  be physically interpreted either in terms of  puffed up branes or in terms of non-singular bubbled geometries with cohomological fluxes. The prototypical examples include the holographic RG flows to  $\Neql1$ supersymmeric gauge theories in four dimensions \cite{Polchinski:2000uf, Klebanov:2000nc, Klebanov:2000hb}, the description of universal effects of  confinement  in $\Neql1$ supersymmeric gauge theories in four dimensions \cite{Gopakumar:1998ki,Dijkgraaf:2002fc,Dijkgraaf:2002vw,Dijkgraaf:2002dh}  and the description of droplets in free-fermion models  \cite{Lin:2004nb}.

The central issue in all of these models is to find the correct holographic realization of the infra-red phase or phases of the dual theory.  For $\Neql1$ supersymmetric gauge theories in four dimensions, this means identifying the correct vacuum or vacua, determining the holographic description of the order parameter of the new phase, like the gaugino condensate, and finding, within supergravity, the dynamically generated scales of chiral symmetry breaking and confinement.  In this context, the ``wrong answers'' were extremely instructive in highlighting the features of the ``right answers.''  For example, the RG flow solution in  \cite{Klebanov:2000nc}  was a crucial step in finding the correct physical answer  \cite{Klebanov:2000hb} but the former flow was too simple and did not turn on fields that were dual to the gaugino condensates.  Similarly, simple attempts \cite{Girardello:1999bd,Pilch:2000fu} to study holographic duals of the $\Neql1^*$  vacua  resulting from turning on masses in $\Neql4$ gauge theories were much too  symmetric and did not involve enough  fields and dual operator vev's to capture the correct holographic infra-red limit given in \cite{Polchinski:2000uf}.

In all these instances, the wrong infra-red phase was singular and  either involved too much symmetry or too few fields with too few fluxes to capture the correct physics.  On the other hand, the correct infra-red physics involved {\it smooth geometries} with precise holographic meaning and in which new homology cycles are blown up and threaded by cohomological fluxes\footnote{As we will explain in Section \ref{Sect:Mechanism}, the resolution of a certain singularity via brane polarization or via smooth cohomology fluxes reflects essentially the same underlying mechanism.}. These fluxes were dual to order parameters of the new IR phase and the scale of the bubbles became the dynamically generated scale of the new phase.

The issue with black-hole microstates is very similar.  There is a CFT at infinity and for each microstate of this CFT we expect that certain operators will have non-trivial vev's.  The problem then becomes one of finding the ``correct'' infra-red phase of the object.  One possible answer is, of course, the black hole, but this is wrong in that it leads to all the vev's of the boundary theory being zero, as well as to a continuum of excitation energies.  As in the holographic description of $\Neql1$ supersymmetric gauge theories and of the microstates of the D1-D5 system \cite{Lunin:2001fv, Lunin:2002iz, Kanitscheider:2007wq},  the ``correct bulk phase'' dual to a certain microstate should involve turning on a rich variety of supergravity modes and their dual vev's and flowing to a smooth geometry with a clear holographic interpretation. This is precisely what  is done by microstate geometries and it seems bizarre and arbitrary to adopt one set of rules for the holographic dual of field theories and another set of rules for the holographic dual of a black-hole CFT.  Such an unnatural logical split also seems especially bizarre given that the idea that black-hole evolution should be unitary got one of its strongest boosts 15 years ago precisely because of  holographic duality and the AdS/CFT correspondence.

We therefore take the only reasonable and consistent position:   Microstate solutions describe the correct set of IR phases of the black-hole CFT.  There are some  important lessons to be drawn from this:
\begin{itemize}
\item The transition scale,  $\lambda_T$, is the analogue of the chiral symmetry breaking scale. It is a new parameter in nature that sets the scale of the infra-red physics.   Classically it is  freely-choosable  but in the quantum theory it is  a dynamically generated scale arising from the strongly coupled theory underlying the black hole. Different black holes may have different IR phases that dominate their structure and thus have different transition scales.
\item  The energy gap, $E_{gap}$, is the energy scale of microstate {\it fluctuations} and is physically very different from the transition scale.  The transition scale sets the size of the homology cycles and the scale of the new phase while the gap relates to the fluctuation spectrum of and around those cycles and within the new phase.
\item  The  infra-red physics is thus a combination of the transition to the infra-red phase and the formation of the large-scale geometry and then the detailed fluctuations of that geometry.   These two phenomena are independent and can be addressed separately.
\end{itemize}

With this in mind, one can now interpret the study of multi-center black-holes using quiver quantum mechanics \cite{Denef:2000nb,Denef:2002ru,Denef:2007vg} or black-hole deconstruction \cite{Denef:2007yt,Gimon:2007mha} as very much in the spirit of the fuzzball and  microstate geometries programme\footnote{The early work on quiver quantum mechanics and multi-center four-dimensional black holes \cite{Denef:2000nb,Denef:2002ru,Bates:2003vx} predates fuzzballs and microstate geometries, but because all four-dimensional solutions either have horizons or naked singularities, its relation to  fuzzballs and microstate geometries was only understood when it was realized that certain of these singular solutions become smooth horizonless bubbling microstate geometries when uplifted to five or six dimensions \cite{Balasubramanian:2006gi, Bena:2008dw}.}.
 The splitting of a black hole into many centers is exactly the four-dimensional analogue of the geometric transition we have discussed, and from a holographic perspective the presence of structure in the transverse space that this splitting achieves appears to be imperative if one is to construct the holographic dual of a pure state of the boundary theory.

Once one has understood that a black-hole microstate geometry with non-trivial size in the transverse directions is dual to an infra-red phase of the dual boundary theory, one can ask the (secondary) question of the extent to which classical supergravity solutions can describe the quantum fluctuations of the boundary theory and sample microstates of the black hole with sufficient density so as to build up a reasonable semi-classical description of the microstates and their thermodynamics.  We would like to stress again that for both BPS and non-extremal black holes this question is distinct from the geometric transition and the identification of the infra-red phases that it represents.  While we are optimistic about the role of microstate solutions, one may ultimately find that classical supergravity solutions do not capture sufficiently many of the microstate degrees of freedom but even then the microstate geometries remain of critical importance because of the infra-red phase they represent and the mechanism they provide for holding up the corresponding ``topological star.''

The fact that every bubbled microstate geometry represents a different infra-red phase described by a different effective theory, means that there is a whole ``Landscape'' of black-hole infra-red theories and that black-hole microstates can differ in terms of not only the microscopic fluctuation structure but also on the larger-scale ($\lambda_T$) of these effective theories at the bottom of a particular black hole.  This generates a very interesting set of questions about whether some parts of this Landscape are preferred over others. 

Another lesson that we would like to recall from quantum field theory is that one of the primary reasons for studying $\Neql1$ supersymmetric gauge theories, whether or not they are realized in nature, is that such theories have enough supersymmetry to help with the analysis and yet sufficiently little to enable confinement and chiral symmetry breaking.  They thus make excellent models for understanding these phenomena with the hope that the supersymmetric versions are not very different from their real-world counterparts. It is in precisely this spirit that we believe that the properties of microstate geometries that we are highlighting here are those  universal features that will generalize from  BPS to non-BPS, non-extremal black holes.

\subsection{Entropy and fluctuating microstate geometries}

Having given us a much better and deeper understanding of the infra-red ground states of the black-hole field theory, microstate geometries can also capture, semi-classically, a vast amount of the black-hole microstate structure. The bubbling solutions that are simplest to construct have a four-dimensional Gibbons-Hawking base space, and are in one-to-one correspondence with multi-center four-dimensional solution of the type \cite{Denef:2000nb,Denef:2002ru,Bates:2003vx}. However, it is clear both from the fact that they are more symmetric than the typical CFT states and from an explicit counting \cite{deBoer:2009un}, that the entropy of these solutions will not suffice to account for the entropy of the dual black hole. 

To obtain solutions that are dual to more typical states of the CFT one has to consider geometries that are less symmetric, and hence do not have a four-dimensional quiver-quantum mechanics interpretation. We have been so far able to obtain two kinds of such solutions: The first are five-dimensional solutions obtained by placing BPS supertubes \cite{Mateos:2001qs,Lunin:2001fv,Emparan:2001ux,Lunin:2001jy,Lunin:2002iz} into BPS geometries. Supertubes are parametrized by several continuous functions of one variable, and their shape modes translate, in the fully back-reacted geometry, into shape modes for the cycles of the bubbling solution. The naive expectation is that these shape modes can provide an entropy that grows as $S \sim Q$ but a novel phenomenon, called {\it entropy enhancement}, was discovered for such fluctuations deep in a microstate geometry \cite{Bena:2008nh} and one can argue that such states could account for an entropy that grows as $S \sim Q^{5/4}$.  While this constitutes a vast number of BPS microstates, the fact that this falls short of $Q^{3/2}$  means that these solutions do not capture sufficiently many degrees of freedom to yield a semi-classical description of  black-hole entropy.

The humble supertube is a beautiful and simple bound state of two electric charges and one dipole charge and its shape modes can account for the entropy of the two charge-system but it will literally not stretch to the description of the entropy of the macroscopic  three-charge BPS black hole.  However, it has been conjectured \cite{Bena:2011uw} that there is a new bound state, called {\it the superstratum}, made of three electric charges and three independent dipole charges that could achieve this goal. This superstratum should source a smooth BPS microstate geometry in six dimensions that is parameterized by four functions of {\it two variables}. There are two arguments for its existence.  The original one  \cite{Bena:2011uw} was based on a double supertube transition \cite{deBoer:2010ud}, and showed that in six dimensions there exist BPS bound states of strings, branes, momentum and KK-monopoles that have the charges of the D1-D5-P black hole and preserve the same four supersymmetries irrespective of two orientations. This strongly suggests that these bound states could be patched up into a BPS configuration whose shape is parameterized by four arbitrary {\it functions of two variables}.  The second argument uses a closed string emission calculation in D1-D5-P microstates to argue that the fields sourced by these microstates depend on functions of two variables \cite{Giusto:2012jx}. One can also argue using a MSW-like argument \cite{Maldacena:1997de} that such objects -- if they exist -- will carry an entropy that exactly matches that of the dual black hole \cite{BSW}.  

The conjectured form of the superstratum \cite{Bena:2011uw}  outlines an algorithm through which  it might be constructed in supergravity.  The computation is extremely challenging but the discovery of the  linear form of most of the underlying BPS system \cite{Bena:2011dd} may have brought the explicit construction within reach.   So far, partial progress in this direction has led to new families of  BPS supergravity solitons \cite{Niehoff:2012wu}, called superthreads and supersheets,  that carry three charges, two dipole charges and fluctuate as  functions of one variable (superthreads) or two variables (supersheets).  These solutions still have singular sources because they are missing one dipole charge that, if it could be incorporated, would yield the smooth multiple superstata.  There is also a class of  microstate geometries that  come very close to obtaining a full superstratum \cite{Niehoff:2013kia} in that they are completely smooth and fluctuate as a function of two variables but the fluctuation functions are still sums of different functions of one variable. 

One can also try to obtain a superstratum solution by perturbing around a round supertube solution, and showing that the perturbation spectrum is parameterized by functions of two variables. So far the solutions obtained in this way are not smooth \cite{Shigemori:2013lta}, but we hope that by marrying the methods of \cite{Niehoff:2013kia} and \cite{Shigemori:2013lta} we will be able to obtain smooth solutions that depend on functions of two variables, and thus prove that superstrata exist.

The fact that such a smooth, solitonic bound state can emerge only for the three-charge system, which is precisely where a black-hole horizon becomes macroscopic, and that the entropy of this object could match exactly the entropy of this black hole is in our opinion very suggestive that we are on the right track for proving that the microstates of three-charge black holes are described by microstate solutions.

Of course, once one constructs a family of fully back-reacted solutions that depend on functions of one or two variables, these solutions are part of a moduli space that classically has infinite dimension. To find how many distinct microstates one has, one needs to quantize this moduli space, and this has been done both for supertubes in flat space \cite{Palmer:2004gu,Rychkov:2005ji} and in deep scaling solution \cite{Bena:2008nh,Bena:2008dw}. The ``quantum microstate geometries'' we obtain from this quantization correspond to coherent states in the dual CFT. It is known that coherent states can give an overcomplete basis for the Hilbert space, but the question that one can ask then is how much of this Hilbert space will be spanned by the microstates that correspond to smooth microstate geometries, how much will be spanned by the microstates dual to ``microstate solutions'' that are not smooth in supergravity but have a string theory or a D-brane interpretation, and how much will spanned by wildly fluctuating quantum fuzz that has no interpretation in terms of anything we can identify as weak-coupling strings, branes or geometry. 

Clearly, to find the answer to this question one needs to construct and count all microstate geometries and microstate solutions, find what are the typical states, and figure out how close these states are from being described using supergravity. This is exactly what has been done for the D1-D5 system, where it was found that typical states are right at the border of validity of supergravity \cite{Lunin:2002iz}. This result came about via some rather non-trivial physics, and it is, in our opinion, premature to venture any speculations as to what the situation will be for the D1-D5-P microstates before one constructs and counts them. The physics of these microstates is clearly richer than that of D1-D5 microstates: for example adding momentum can turn singular D1-D5 throats into thick black-hole-like throats and has the potential to also transform certain singular D1-D5 solutions into smooth solutions. Hence, it may well be possible that the typical D1-D5-P black-hole microstates will be better behaved than the microstate geometries of the D1-D5 system, and therefore the typical microstates of the D1-D5-P black hole will be described by smooth low-curvature supergravity solutions\footnote{In \cite{deBoer:2009un} it has been argued that supergravity will not have enough states to account for the entropy of the black hole. These arguments were based by two counting calculations that consider configurations less general than the ones considered here. The first is a multi-center four-dimensional quiver, which when uplifted to six dimensions has two extra $U(1)$'s compared to our solutions. The second is a graviton gas in $AdS_3$, whose back-reaction does not capture the non-trivial topology of the bubbling geometries. From the perspective of the graviton gas these geometries would correspond to non-perturbative strong coupling degrees of freedom, and the weak-coupling counting in \cite{deBoer:2009un} does not see them, much like calculations done with the Yang Mills Lagrangian do not see confinement.}.

If the final answer to the question of whether the typical vectors that span the black-hole Hilbert space are dual to smooth microstate geometries is negative, one can still try to argue that these smooth geometries are capturing the essential physics of the typical microstate solutions, and that they give a good sampling of this Hilbert space. Indeed, in order to obtain a semi-classical description of the thermodynamics of a given system one only needs to sample its Hilbert space at a very sparse rate. It is worth remembering that the kinetic theory of gases was developed long before quantum mechanics and it only required the quantization of phase space and not the value of $\hbar$.    Put differently, one can get a good description of an ideal gas in a room if the atoms are treated as hard spheres and have a size that is thousands of times larger than they are in reality. Similarly, one might hope to get a useful description of black-hole thermodynamics from supergravity modes that lie within the validity of the supergravity approximation even it the typical black-hole microstate solutions are not describable by supergravity.

\section{One Mechanism and its Three Hypostases}
\label{Sect:Mechanism}

Having described in detail the physics of smooth microstate geometries and the scales associated to them, we have seen that the only way a microstate geometry of a BPS or a non-BPS black hole can be stationary is if it has multiple cycles supported by cohomological flux. These geometries do not collapse into a black hole essentially because in that limit the cycles become very small, and the energy of the fluxes diverges. 

It is important to understand what this mechanism becomes in various regions of the parameter space where microstate geometries  become microstate solutions or even more generic fuzzball states. More precisely, such geometries can have singular limits that make sense within string theory but cannot be properly described by supergravity.  As we will explain, the microstate geometries and the mechanism that supports them can transform into other familiar stringy phenomena that can support microstate solutions and fuzzball states. It is in this sense that we mean that there is really only one mechanism and that it can have (at least) three avatars.

One class of singular supergravity limits can occur when, by decreasing $g_s$, the size of the bubbles, $\lambda_T$, becomes of order the Planck scale in some region of the solution. To understand this limit one should recall that in general bubbling geometries when a the scale of a cycle becomes Planckian, one gets an ``inverse geometric transition,'' in which the cycle is replaced by a brane. After all, the way the first bubbling geometries were obtained \cite{Bena:2005va} was through a geometric transition from a singular brane source to a smooth solution. 

One must also replace a smooth cohomological cycle with a singular brane source when changing between certain duality frames. Recall that the two-charge supertube solution is smooth in the D1-D5 duality frame, but becomes singular in other frames, such as the D0-D4 frame. Hence, a smooth black-hole microstate geometry in one frame can be transformed into a solution with a singular brane source. 

The singular brane sources obtained by dualizing smooth solutions have a common characteristic: they have locally 16 supersymmetries, and have non-trivial dipole charges. Hence, one can think about these branes as coming from the puffing up of the branes corresponding to the black-hole charge and converting them  into configurations with a dipole moment.  There are several process that can cause such branes to puff up: they may be forced to do so by external fields (as it happens in the Myers effect \cite{Myers:1999ps}), or under the influence of angular momentum (as it happens to supertubes \cite{Mateos:2001qs}). In this limit, the mechanism that was responsible for keeping the microstate geometry from collapsing, bubbling, becomes the mechanism that prevents the collapse of the puffed-up branes: local angular momentum for supertubes, external fields for non-rotating dipole branes, and possibly a combination of the two for more generic configurations. Again, we do not expect non-extremality to destabilize these configurations: it is known for example that adding non-extremality to a black ring simply adds another term in the balance between the centrifugal force and the gravitational attraction that gives its radius, and makes this radius shrink a little \cite{Elvang:2004xi}. Similarly, when warming up Polchinski-Strassler \cite{Freedman:2000xb} one still finds minima that contain polarized branes, at least for temperatures smaller than the energy scale of the polarization potential. 

In addition to brane polarization, or puffing up, the mechanism for upholding the black-hole microstate solutions can have a third incarnation.  In the limit when the back-reaction of the branes that puff up becomes small the microstate solutions can be supported by non-Abelian effects. In this regime both supertubes and polarized branes can be described using the non-Abelian field theory that describe the polarizing branes \cite{Myers:1999ps,Bena:2001wp,Bak:2001kq}. The scalars that live on these branes, or the bi-fundamentals corresponding to strings stretched between different types of branes, acquire vacuum expectation values that do not commute, and it is these non-commuting degrees of freedom that end up describing the puffed-up configuration. A similar phenomenon can be observed in the quiver quantum mechanics description of multi-center configurations, where in certain Higgs-branch states the bi-fundamentals acquire non-commuting vacuum expectation values, which then have an interpretation in terms of branes with non-trivial dipole moments\footnote{These Quiver Quantum Mechanics configurations oftentimes go under the name of ``Coulomb-branch states'', they should not be confused to the ``Coulomb branch'' solutions which correspond to unbound configurations of branes that can move freely on the Coulomb branch. As argued above, the latter do not have the right properties to correspond to black-hole microstates (they do not grow with $g_s$), while the former do.}.  

Again, these non-Abelian vacua are not expected to be destabilized by non-extremality, and hence will describe certain regions of a black-hole microstate at least some finite distance away from extremality. Furthermore, there is evidence that non-Abelian physics can affect the horizon scale even far away from extremality: a recent study of thermal configurations in the BFSS matrix model \cite{Berenstein:2013tya}, has found that the effective field theory of a probe breaks down at a particular location that corresponds to the black-hole horizon. 

One can also imagine a very large microstate geometry that is everywhere smooth except in a small spatial region, where the physics is described by the non-Abelian degrees of freedom of certain branes.  This solution exists in the same region of moduli space as the classical black hole, and as one increases the coupling constant one can make this non-Abelian degrees of freedom ``condense'' into a polarized D-brane or into a supertube, and by further increasing the string coupling one can transform these polarized branes via a geometric transition into a smooth geometry. Clearly, in all three regimes of parameters the solution should be accepted as physical, despite the lack of a smooth supergravity description in the first two.  

It is for this reason  that we want to think of cohomological fluxes, brane polarization and non-Abelian effects as three hypostases of the one underlying mechanism and that each apparently distinct phenomenon is simply the manifestation of this mechanism is different limits of the theory. The result of \cite{Gibbons:2013tqa} suggests that if any time-independent support mechanism can be given a supergravity limit then it must correspond to bubbles being held up by flux. Conversely, we believe that any viable means of support for a fuzzball must be some limit of the cohomological flux mechanism, and hence if one could, for example, compute the full back-reaction of coherent matrix model states of type analyzed in \cite{Berenstein:2013tya}, they will become a solution of the type we discuss here. 

Of course, one can also imagine a microstate solution where most of the bubbles become very small, and most of the configuration is described by non-Abelian physics, or other solutions where all three hypostases of this mechanism are valid descriptions of the physics in different regions. Clearly, some of these solutions will be more typical than others, and, as we explained in the previous section, at this point we do not know what will be the features of the typical solutions, and which of the three incarnations of this mechanism will describe them. However, we are confident we have identified the underlying mechanism for preventing a horizon-sized configuration from collapsing, and we believe that this mechanism will be robust when moving away from extremality.

\section{Near-BPS Configurations}
\label{Sect:NearBPS}
The work on microstate geometries over the last decade seems to have given us a relatively good picture of BPS microstate geometries.  There has also been an extensive amount of work on non-BPS, extremal solutions but there are still only a handful of explicitly-known non-extremal microstate geometries. Constructing such non-extremal configurations is, of course, crucial to establishing the complete fuzzball proposal. 

Indeed, the physics of the known microstate solutions of {\em extremal} black holes indicates very strongly that the region behind the horizon of these black holes is wiped out by structure arising from the resolution of the singularity, and therefore  there is no space-time corresponding to what might have have been the black-hole interior. From the perspective of many a general relativist, this is not  unexpected. The inner horizons of non-extremal black holes are known to be unstable to classical fluctuations, and so are the horizons of extremal black holes \cite{Penrose:1968ar, Brady:1995ni, Dafermos:2003wr, Poisson:1990eh,Marolf:2010nd}. This can also be seen by realizing that extremal four-dimensional black holes have an infinite $AdS_2 \times S^2$ throat, and this throat does not support finite-size energy perturbations  \cite{Maldacena:1998uz}. From this perspective, the extremal microstate solutions that we have been constructing are simply end-points of the instability of extremal horizons, and thus illustrate the mechanism by which the singularity gets resolved and wipes out the region between it and the inner horizon. However, if one tries now to extend this point of view to non-extremal black holes, it may well be that the resolution mechanism via microstate geometries only affects the inner horizon, while leaving intact the region between the inner and outer horizon. 

Nevertheless, this is not what the fuzzball proposal proposes. In order to recover information from the black hole, the region between the inner and the outer horizon, which is in the causal past of the singularity, must also get wiped out by fluctuations, and there should be no spacetime inside that region either. The only way to show that this indeed happens is to take the known extremal microstate geometries away from extremality and argue that they do not develop a horizon in the process, and more generally to try to construct as many non-extremal microstate solutions as possible.

There are only two known families of such solutions, and these go under the names of JMaRT \cite{Jejjala:2005yu} and running-Bolt solutions \cite{Bena:2009qv,Bobev:2009kn}. These solutions are unstable \cite{Cardoso:2005gj,Avery}, and for the JMaRT solutions there is a very beautiful story that relates their instability to that of the dual boundary state \cite{Chowdhury:2007jx,Avery:2009tu,Avery:2009xr}, which in our opinion constitutes a striking confirmation that the fuzzball proposal applies to non-extremal black holes. It is important to understand whether the instability of the JMaRT and running-Bolt solutions is only a peculiar feature of these solutions, or is a generic feature of all non-extremal microstate geometries \cite{Chowdhury:2008uj}.  More generally,  it is obviously of considerable interest to try to build up intuition about generic non-extremal microstate geometries by constructing families of examples.

As we remarked in Section \ref{Sect:topology}, one might be able to make some classes of non-extremal microstate geometries using inverse scattering methods and we suspect that this might be the most direct method of generalizing the  JMaRT solution. Here, however, we wish to examine another natural starting point for studying generic microstate geometries:  the construction of {\it near-extremal} microstate geometries.   There are several approaches to this problem and they can be roughly characterized as perturbative methods, global supersymmetry breaking and non-perturbative techniques.

\subsection{Perturbative techniques}
\label{Sect:Perts}
 
One of the simplest ways to move away from extremality is to consider fluctuations around a supersymmetric solution.   For example, demanding that a state carrying momentum on a string or D1-brane be BPS usually requires the momentum to be purely left-moving or right-moving and so adding small amounts of modes with the ``wrong'' motion is a simple way to break supersymmetry.  More generally, any brane can be given non-BPS fluctuations as well as BPS modes, and in general understanding the non-BPS modes is crucial if one wants to quantize correctly the BPS modes \cite{Palmer:2004gu}.  In terms of microstate geometries such fluctuations correspond to shape modes on the homology cycles.  There are sometimes ways to make shape modes that are BPS but there are generically many more modes that are non-BPS and move the entire solution away from extremality.  

In the same spirit, one can also generate near-BPS, time-dependent solutions by allowing slow motion on the moduli space that characterizes a given BPS solution. If one has a BPS configuration that exists because of the simple force balance of electrostatics and gravity then this moduli space can be non-compact and the centers of the solution can fly apart or have some unusual coincidence limits (see, for example, \cite{Michelson:1999zf,Michelson:1999dx}).  
However,  BPS microstate geometries are different: they have non-zero magnetic fluxes threading certain cycles, as well as non-trivial angular momentum that depends upon the layout and separations of the cycles. These fluxes can act to bind bubbles together and compactlfy the moduli space. Indeed, {\it scaling}  microstate geometries have all the non-trivial cycles bound together at the bottom of a long AdS throat and so these configurations cannot fly apart.  On the other hand, these scaling solutions have coincidence limits, which corresponds to the throat of the solution becoming larger and larger while the cap remains self-similar. Classically, these coincidence limits appear to be non-compact sections of the moduli space but, as we will discuss in Section \ref{Sect:Quantum}, semi-classical quantization shows that these classically non-compact limits have finite volume in phase space and so the moduli space is effectively compact. 

There are several ways in which one may hope to study this slow motion on the moduli space. For example, one can examine the first-order back-reaction on the supergravity solution created by this motion, as it was done for in  \cite{Gibbons:1986cp,Michelson:1999dx} for unbound BPS multi-black-hole solutions. Alternatively, one can use the Lagrangian of the Quiver Quantum Mechanics that characterizes these centers at weak coupling \cite{Anninos:2012gk}, and to hope that the near-BPS  physics captured by this Lagrangian is robust as one goes to the regime of parameters where the centers back-react.  

The important point we would like to stress is that fluctuations and motion on the moduli space do not change the topology and do not alter the underlying magnetic fluxes.  Thus the transitioned geometry is robust under such perturbations.  This is very different from the intuition that one might have if one insists upon thinking of BPS solutions as supported by a force balance between gravity and electrostatics.  We therefore think that these fluctuation and motion models of near-BPS geometries will provide useful and effective descriptions of black-hole microstates that have a very small temperature.  Indeed, this perspective very much informs our much more speculative discussion in Section \ref{Sect:Fuzz}.  

Thus the advantage of the perturbative approach is that it enables some simple gedanken experiments that give insight into what may occur when one warms up a BPS fuzzball. The disadvantage is that of any perturbative method:  the technical complexity of implementation and the fact that one cannot access the non-perturbative components that will almost certainly be an important part of a generic fuzzball.

\subsection{Global supersymmetry breaking}
\label{Sect:Global}

One of the insights that led to the large new families of non-BPS, extremal solutions and some new non-extremal microstate geometries was that one can assemble configurations whose components are individually supersymmetric and perhaps mutually supersymmetric with other components of the solution but when the configuration is taken as a whole, all supersymmetry is broken (see, for example, \cite{Goldstein:2008fq,Bena:2009ev,Bena:2009en,Bena:2009fi,Bena:2011ca,Vasilakis:2011ki}).  In many of these examples the supersymmetry is broken by holonomy of the gravitational background: the space is populated by supersymmetric combinations of branes and fluxes but the gravitational background does not respect that supersymmetry. 

This approach has, thus far, generated many extremal, non-BPS solutions but we believe that it might be pushed harder and further.
For example, Mathur and Turton \cite{Mathur:2013nja} have recently argued that one might be able to find geometries in which the charge species rotates in such a way that at every point the configuration has BPS structure but infinitesimally neighboring configurations are not quite mutually BPS and this local BPS structure can slowly vary  around the configuration so that widely separated parts could have completely opposite charges. One might even do this in such a manner that the overall configuration is electrically neutral. There is strong evidence that such configurations exist in string theory \cite{BRW}, which might enable one to describe Schwarzschild-like microstate geometries in a manner that locally looks like a perturbation of a BPS structure. 

Another example of a microstate geometry whose supersymmetry could be (partially) broken would be the microstate geometries of supersymmetric black holes in gauged supergravity, like the Gutowski-Reall black hole \cite{Gutowski:2004ez,Gutowski:2004yv} or its four-dimensional friends \cite{Cacciatori:2009iz,Klemm:2011xw}. If one takes the limit where all the scales are larger than the cosmological constant, these black holes resemble their flat space analogues, and there is a very rich class of bubbling microstate geometries for the latter. If one then imagines slowly restoring the cosmological constant, such that the AdS curvature is  extremely small across the size of the bubbled geometry, there should be no impediment to the existence of  asymptotically-$AdS$ microstate geometries because the geometric transition and the bubbling of microstates is an entirely local phenomenon and the bubbles should not be destabilized by small perturbation. Hence, we believe there should exist very vast families of asymptotically-$AdS$  microstate geometries, and that the reason why they have been not found yet is purely technical: the equations determining all supersymmetric solutions of gauged supergravity \cite{Gutowski:2004ez,Gutowski:2004yv} are non-linear, and even solving them to obtain a cohomogeneity-one solution is a rather technical endeavor.  

It is also possible that there is a class of bubbled microstate geometries whose supersymmetry is completely incompatible with $AdS$ supersymmetry and so such   $AdS$  microstate geometries may necessarily be non-BPS.  This would be consistent with the fact that there are no supersymmetric, $AdS$ black rings \cite{Grover:2013hja}.

\subsection{The general problem: non-perturbative additions}
\label{Sect:General}

As shown in  \cite{Gibbons:2013tqa}, the only way to construct smooth, time-independent microstates for non-extremal, five-dimensional black hole is to arrange cohomological fluxes so that they source opposite charges in different regions. Indeed, this is very similar to the mechanism proposed by \cite{Aganagic:2006ex} to give metastable Calabi-Yau compactifications.  We also believe that the flux-anti-flux solutions will be an essential part of moving beyond near-extremal and closer to Schwarzschild or Kerr microstate geometries.

Building such configurations with full back-reaction is notoriously difficult essentially because blowing up cycles using both fluxes and anti-fluxes is an intrinsically non-perturbative process.  However, in \cite{Bena:2011fc} it was proposed that one could simplify this  problem by considering metastable probe supertubes placed inside a supersymmetric solution and constructing the corresponding near-extremal solutions. These solutions are similar in spirit to the anti-D3 brane probe construction \cite{Kachru:2002gs} of metastable vacua in Klebanov-Strassler \cite{Klebanov:2000hb}, and decay into the supersymmetric vacuum by the same mechanism: brane-flux annihilation. In \cite{Bena:2012zi} this technology was used to construct microstate solutions for near-extremal black holes, and to argue that there exist very large families of such microstate solutions. It is obviously an important problem to construct the fully back-reacted solutions corresponding to these metastable supertubes, and to examine the physics implied by these solutions.   In particular,  it would be very important to see whether this back-reaction maintains the metastability, or creates an ergosphere, or gives rise to other physics. 

As side comment, we should mention that there exists an extensive body of work devoted to understanding the back-reaction of anti-branes in solutions with charges dissolved in fluxes (see for example \cite{Bena:2009xk,Bena:2011wh,Bena:2012bk}), and so far it looks like that this back-reaction may be problematic, and that these anti-branes will not give a viable method to uplift $AdS$ vacua to deSitter \cite{Kachru:2003aw} and to create viable string cosmologies. However,  supertubes have richer charge structure and larger co-dimension (and so there is less likelihood of dangerous logs)  and therefore metastable supertubes could be viable building blocks for non-extremal black-hole microstates.

\section{Lessons from Quantum Mechanics}
\label{Sect:Quantum}

\subsection{Entropy and the transition scale, $\lambda_T$}
\label{EntFavoring}

We have tried to emphasize that, at least in principle, the transition scale of microstate geometries is different from all other scales of black-hole physics. Of course, once one constructs and counts all solutions it may be that the transition scale of the typical microstates will simply be the Planck scale and that all the bubbles will be Planck-scale fuzz stretched across the scale of the would-be horizon. It may also turn out that, on the contrary, a much larger transition scale, $\lambda_T$, will be entropically favored.

Indeed,  large bubbles do not seem to be disfavored. Denef and Moore showed \cite{Denef:2007vg} that the quiver quantum mechanics of a three-node quiver could yield black-hole-like densities of states.  Such configurations correspond to BPS microstate geometries in which there are only three homology cycles each of which has a size that is almost the diameter of the BPS throat.   Thus quantum fluctuations of extremely large bubbles can carry entropy that is close to the required levels in a black hole.

To see why small bubbles may well be entropically disfavored, it is instructive to consider a two-charge microstate geometry that has $N$ supertubes, which give rise, in the D1-D5 duality frame, to $N$ bubbles.  Suppose that we divide each of these supertubes into $k$ supertubes, so that there are $k N$ bubbles.  This means that there are now $k$ times as many moduli and $k$ times as many arbitrary continuous functions that characterize the solution, and hence an entropy gain by a factor of $\sqrt k$ (the number of functions determines the central charge in the effective CFT that gives the entropy). However, we must now partition the fluxes amongst the bubbles so that the overall electric charge remains the same.  Just as in Section \ref{Sect:Typicality}, the energy gap for excitations of  a single bubble or a single supertube is proportional to $\frac{1}{Q^t_1 Q^t_2}$ where $Q^t_1$ and  $Q^t_2$ are the charges of that supertube, so the partitioning increases the energy gap by a factor of $k^2$. Thus, for a total momentum or angular momentum budget, the number of states one can excite goes down by a factor of $k^2$ and so the overall entropy drops by a factor of $\sqrt{k}$. Put simply, making more but smaller bubbles means that there are more things that can fluctuate but it also means that there are many fewer fluctuation modes available to be populated and the latter effect seems to dominate\footnote{We are grateful to Slava Ryckhov for this argument.}.

There is another argument that small bubbles will not be able to carry a black-hole-like entropy which is based on the computation of the index of multi-center solutions in four-dimensional $N=4$ supergravity and an argument that such states do not contribute to the index of the black hole \cite{Dabholkar:2009dq}. It is unclear to us  whether this argument applies to the fluctuating bubbling geometries that we discuss here, which do not descend to four-dimensional supergravity, and how this argument can be reconciled with the fact that solutions with long throats have exactly the correct mass gap to belong to the typical sector of the CFT\footnote{A recent discussion on this issue has appeared in \cite{Chowdhury:2013pqa}.}. However, if this argument applies to fluctuating bubbling solutions it would indicate that the place to look for the black-hole entropy is not multiple bubbles but a single large bubble. The superstratum entropy calculations of \cite{BSW} and the Higgs-Coulomb computations of \cite{Bena:2012hf,Lee:2012sc} also appear to point in the same direction. 

\subsection{Infalling shells and the formation of microstate geometries}

In the fuzzball program one wants to replace classical black holes with smooth, horizonless geometries populated by an extremely dense microstate structure. However, it seems {\it almost} inconceivable that this could happen in a dynamical process. Suppose one sets up a spherical shell of dust of ten solar masses and with an initial radius on 10,000 AU:  It seems impossible that this could collapse into anything other than the standard Schwarzschild black hole. On the other hand, we have been arguing that black-hole physics is described by a new phase of matter and that the phase transition is described by a geometric transition to some kind of topological star.  

What makes the black-hole phase transition so radically different from our normal experience of macroscopic phase transitions is the staggering density of states in a black hole.  Mathur  has argued  \cite{Mathur:2008kg,Mathur:2009zs,Mathur:2013gua} that the last stages of a spherical infall are best described by quantum tunneling transition rather that a simple classical evolution in General Relativity.  The argument is simple: the amplitude for a particle to tunnel from outside the black hole into a fuzzball state inside where the horizon would be is extremely small:  of the order $ e^{- \alpha M^2}$, where $M$ is the mass of the matter and $\alpha$ is a number of order unity.  On the other hand, the number of states in the fuzzball is given by a number of order $ e^{+  M^2}$ and so by Fermi's Golden Rule the probability of tunneling is of order $1$.  

This seems surreal because it implies that macroscopic pieces of space-time, in which the curvatures are small and thus supergravity should apply,  are being wiped out by quantum fluctuations.  Obviously, such a complicated dynamical transition is very hard, if not impossible to compute in any detail. However, a similar phenomenon has been seen and computed in time-independent microstate geometries. Recall the discussion in Section \ref{Sect:Typicality} where the quantization of BPS geometries could wipe out macroscopic regions of space time in which curvatures were low and the supergravity approximation should be reliable. While this is not a dynamical process it does exemplify the kind of quantum transition to a fuzzball state advocated in  \cite{Mathur:2008kg,Mathur:2009zs,Mathur:2013gua}.

The central idea of this particular black-hole paradigm is that the staggeringly high density of states of the black hole means that quantum effects involving the black hole can change space-time geometry on macroscopic scales. It might seem that supergravity can tell us very little about such a regime, however this may be too naive.  Just as one learns that a diatomic molecule can, in some approximation, vibrate like a classical harmonic oscillator and that these modes admit a simple quantum description, so might one hope that naive semi-classical quantization of supergravity moduli spaces could give real, physical models of some black-hole structure.  The classical statement that the supergravity approximation should hold simply means that higher-mass, stringy and Kaluza-Klein fluctuations effectively decouple and that supergravity should capture the dominant dynamics.  One can then apply semi-classical quantization methods, such as those in  \cite{deBoer:2008zn,deBoer:2009un} or maybe use instantons and WKB methods to estimate tunneling amplitudes, and perhaps derive some useful approximate picture of the state, and even the evolution of states, in a black hole.  Thus, even though the classical supergravity description would not apply to the typical black-hole microstates, there could well be a very important role for some simple and naive quantization of the microstate geometries described by supergravity.

\section{Schwarzschild and Kerr microstates - some wild extrapolations}
\label{SKmicrostates}

We now try to put together a ``best guess'' of what Schwarzschild or Kerr microstates might look like.  We should stress that we are making wild extrapolations based upon extremal and BPS data and drawing extensively on the intuition about near-BPS solutions described in Section \ref{Sect:NearBPS}. Our viewpoint is based more upon what we hope may emerge from such extrapolations rather than upon detailed calculations.  However, by the present standards of this field, we think that our ideas, wild as they may be, are worthy of consideration and perhaps have a marginally firmer grounding in mathematics, if not in reality, than some other proposals.

\subsection{A Fuzzball perspective}
\label{Sect:Fuzz}

First, the correct description of a black hole will have neither a singularity nor a horizon\footnote{Obviously the Schwarzschild and Kerr solutions exist, but they are far too symmetric and, according to the fuzzball paradigm, are not the correct description of physics at the horizon any more that a delta function can really replace the details of a planet as a source of a gravitational field.}.  The supergravity description ``caps off'' the geometry before a horizon forms and replaces the collapsing matter with a ``topological star'' whose surface area is of the same order as, but larger than,  that of the would-be black hole.  The mechanism to support this ``topological phase''  of the star necessarily involves the opening up of extra dimensions and the geometric transition in which the majority of the perturbative matter is replaced by solitonic configurations with topological fluxes.  The transition to this state of matter requires non-trivial Chern-Simons interactions and the opening up of non-trivial  topology in the space-time.  The long-term (classical) stability of black holes, and their slow decay via Hawking radiation, requires the existence of metastable flux-anti-flux configurations in the topological star.  Given that such meta-stable flux configurations are the basis of a substantial number of stringy cosmologies, it may be that stabilizing a black hole is not such a tall order compared to stabilizing the entire universe.

The geometric transition to a topological phase thus provides the structure and mechanism to support the fluctuations (classical, quantum and stringy) that describe the microstates of the black hole.  As we explained in Section \ref{Sect:Holography}, associated with this are two new very important physical parameters: The transition scale and the gap energy.  The former sets the scale of the geometry of the transition and the latter determines the spectrum of fluctuations.  The fluctuation spectrum of a topological star is thus discrete.  In particular, the transition and the gap energy reflect the fact that the topological star has a finite ``depth'' in that there is a maximal (but still very large) redshift, $z_{max}$, in traveling from the center of the star to infinity.

The topological star will slowly decay.  The non-BPS fluctuations can simply radiate away as stringy modes can be emitted from branes and other strings. There is also the possibility of  brane-flux annihilation which in the large-flux limit becomes flux-anti-flux annihilation.  Such events might have a different energy signature from the simple, perturbative radiation of stringy states.  
All of these decay channels might be relatively rapid in the local frame in the heart of the topological star, but the  very large finite red-shift will mean that these  decays will be very slow as seen from infinity.  However, one hopes that radiation from the decay of typical microstates would look to first approximation like Hawking radiation, but will manifestly encode all the details of the microstate structure; hence, the evaporation of the ``topological star'' will, from the information-theoretic perspective, be no more mysterious than burning a piece of coal. 

The temperature of the topological star or fuzzball as seen from infinity will be the same as the Hawking radiation, but because fuzzballs do not have a horizon and come with two extra scales, the temperature measured at the surface of the fuzzball will not be infinite: it will be of the order of  $T_{Hawking} \times z_{max}$. Discussions about fuzzball complementary aside, it is plausible that the same temperature will be measured by an infalling observer. Hence the fuzzball state and the firewall have some level of commonality:  The fuzzball is hot, but unlike the firewall, not infinitely so.

Once a fuzzball has settled into such quasi-equilibrium with its slow decay through  Hawking radiation one expects that all the microstate structure should, in principle, be accessible from outside.  One way to access this is by analyzing the multipole expansion of correlation functions functions near the black hole.   For a classical black hole there will be no structure in this expansion because a black hole has no hair.  

For thermal correlators near a heat bath one expects an exponential fall-off of correlation functions with a relatively short correlation length that depends upon the temperature. This is precisely what one should expect from the black-hole microstate structure with correlation length most likely related to the transition scale.  One does not see such exponential decay of correlators in individual BPS microstate geometries:  such correlation functions have multipole expansions with power-law decay. This is no more surprising than saying that an individual Ising configuration or the  individual states of molecules in a box of gas have a non-trivial multipole expansion.

The exponential decay of the correlation functions only arises from thermodynamic or quantum averaging. Indeed, microstate geometries will form a coherent-state basis of the Hilbert space of a certain black hole, but that a typical state of this system will be in general some random vector in the Hilbert space that does not have a geometric meaning, but rather is a quantum superpositions of many geometries. To compute the vev's of such a Haar-typical state, $|\psi\!>$, one has to compute the vev's in each of the $e^S$ microstate geometries appearing in the expansion of $|\psi\!>$, and superpose them. The resulting average will suppress the multipole details of each state by $e^{-S/2}$, the usual statistical square root of the number of states. Hence, we expect that averaging over the multitudes of BPS states will give correlators an exponential decay,
  
It is unclear what will become of the bubbles and fluxes once the thermal fluctuations of the black hole will become larger than the scale of the bubbles, $\lambda_T$. The temperature could clearly collapse the bubbles, but it could also trigger their tunneling into other bubbles, and compensate the rapid  annihilation of flux and anti-flux bubbles by nucleating new ones. This whole picture is based largely upon a combination of very wishful thinking applied to the BPS and near-BPS microstate geometries.  It is therefore important to say something about how one might remotely begin to support, or falsify, the picture outlined above.  

First, we need more examples of Schwarzschild-like or Kerr-like microstate geometries, supported by flux and anti-flux.  This is, a priori, very hard because one is dealing with the full non-linear equations of motion.  On the other hand, and as we remarked earlier, there may be examples of such microstate geometries with enough symmetry that they could be obtained by inverse scattering methods or numerics.

One of the surprises over the last few years is how BPS techniques and equations could be adapted to the investigation of extremal, non-BPS solutions. It might be hoped that extremal Kerr could be ultimately be shown to have some underlying simplification in the structure of the equations that govern them. If this could be achieved, then the near-Kerr objects and  near-Kerr microstate geometries will be of immediate relevance to astrophysics. Such a structure is revealed for example when one tries to embed the five-dimensional near-horizon, extremal Kerr geometries in String Theory \cite{Bena:2012wc}. The geometries one obtains allow for exactly the same type of multi-center bubbling one obtains in BPS and almost-BPS solutions, and constructing microstates geometries for such black holes is currently under consideration \cite{NHEK-String}.

\subsection{Firewalls}
\label{Sect:Fire}
\vskip -0.1cm
{{\small  Arthur: { \it ``Ah. Look, the statue, how did you get the cup bit to stay where it is unsupported?'' }}}
\newline
 {\qquad   {\small  Wise old bird:  {\it``It stays there because itÕs artistically right.''}}}
\newline
 {\qquad {\small  Arthur: { \it ``What?'' }}}
\newline
 {\qquad {\small Wise old bird: { \it ``The law of gravity isn't as indiscriminate as people often think. You learn things like that when you're a bird.''} }}
\newline
\rightline{\small  Douglas Adams,  {\it    The Hitch-Hiker's Guide to the Galaxy,} }
\smallskip

It would be remiss if we did not make some remarks about firewalls \cite{Almheiri:2012rt,Mathur:2012jk,Susskind:2012rm,Bena:2012zi,Susskind:2012uw,Avery:2012tf,Almheiri:2013hfa,Avery:2013exa,Verlinde:2013uja,Maldacena:2013xja,Mathur:2013gua}. Reduced to its most basic form, the idea is to start with Mathur's quantum information argument \cite{Mathur:2009hf,Mathur:2012np,Mathur:2012dxa} that shows if information is to be recovered from a black hole then there must be corrections of $\cO(1)$ to the Hawking states at the horizon  and then add the infinite blueshift associated with falling into a horizon. This means that a freely falling  observer  will encounter quanta of arbitrarily high energy near the horizon and will thus burn up\footnote{The AMPS paper \cite{Almheiri:2012rt} also contains a  detailed discussion of black-hole complementarity  but we will not address this issue here.}.    

To solve the problem of information loss and to create a firewall one needs erect a lot of new structure at the horizon scale. Unfortunately, like much of the discussion of the information problem over the last three decades, a large part of the analysis of firewalls takes place in the framework of four-dimensional relativity and quantum mechanics, sometimes with a pinch of AdS-CFT.  In such a framework it is clear that the no hair theorems guarantee there will not be any mechanism to support structure at the horizon: the firewall will either be expelled or fall into the black hole. Hence, the only way one can hope to support a firewall using general relativity and quantum mechanics alone is to invoking some  {\it Quantum Coyote Principle}\footnote{Gravity does not act until you look down.}.  
 
If one is to construct a mechanism to support a firewall, particularly within some kind of holographic field theory, there are going to have to be some new scales:  one associated with the scale of the supporting structure and another that cuts off the infinite blueshift/redshift at the horizon and introduces a gap into the fluctuation spectrum. This way the horizon structure can be hot, but not infinitely hot.  If one believes that string theory is a viable quantum theory of gravity, then it is the obvious place to look for such a mechanism.  One is then automatically led to supergravity and the sorts of theorems outlined in Section \ref{Sect:topology} and the fact that there must be a geometric transition and a scale associated with it. Thus, at the end of the day, the search for a mechanism to support a firewall will ineluctably lead to microstate geometries.  Hence,  even if your aesthetic principles would prefer something else, you simply have to accept the Holmesian adage: {\it ``When you have eliminated all which is impossible, then whatever remains, however improbable, must be the truth.''}  

Therefore, we believe that the best course to understand what happens to an infalling observer that crosses the black hole horizon is to start with some configuration that actually makes sense within string theory, like a microstate geometry, and analyze the problem from there. The result may either be some from of fuzzball complementarity \cite{Mathur:2010kx,Mathur:2011wg,Mathur:2012jk,Mathur:2012zp}:  certain classes of infalling observers would experience the fuzzballs in much the same way that one would experience geodesic infall into a Schwarzschild black hole, or might be a complete splat, as one expects from firewall arguments and from the original incarnation of the fuzzball proposal (discussed for example in \cite{Bena:2007kg}).  

At this point the physics of the only known large class of non-extremal microstate solutions \cite{Bena:2012zi} appears to indicate very wild fluctuation fluctuations of the force felt by an incoming probe brane, and hence a possible immolation of the latter. However, one must remember that microstate geometries are describing rather coherent states of the system while fuzzball complementarity and the ``empty space-time experience''  require quantum and/or thermal averaging over states.   Thus the microstate geometries suggest a classically violent infall but this may well be an artifact of the basis that has been chosen and soft landings for selected observers may emerge from finding a way to perform quantum superpositions of microstate geometries.

\section{In Lieu of Conclusions}

To a significant degree, Section \ref{SKmicrostates} can stand as our summary and conclusions because our primary goal was to extract universal ideas from BPS and near-BPS microstate geometries and to try to gain some idea of the semi-classical/fuzzball description microstates structure of generic black holes.

Beyond this we would like to observe that, after a decade, the BPS microstate geometry story turned out to be far richer and get much further than any of us, including the authors, ever expected.  In addition, the general view of this program moved from skepticism of the BPS results to a growing acceptance of major aspects of the BPS picture and thence to arguing as to whether it can possibly work for non-BPS or Schwarzschild-like black holes.   Thus the goal posts have  moved from the BPS standard of the 1990's to asking about the full-blooded Schwarzschild result. While this attitudinal shift reflects a skepticism that is the basis of good science, we also find it encouraging in that one begins to feel that the BPS picture is beginning to be properly fleshed out.  On the other hand, asking for Schwarzschild microstates is a very tall demand at present because of the computational complexity of the problem.

One of the other points we want to stress here is that the support of microstate geometries via fluxes is {\it not} simply a finely balanced, purely BPS phenomenon. The topology of microstate geometries is robust under perturbations and so can certainly  give valuable insight into near-BPS solutions.  From this one can then try, as we have here, to get deeper insight into the generic microstate structure for Schwarzschild black holes.  So while the most important problem is obviously the  microstate  structure of astrophysical black holes,  it remains very worthwhile to focus effort and energy on near-BPS microstate geometries.   Such geometries represent a very important middle ground between what is computable and what may yield insight into Schwarzschild microstate structures. 
 
One of the most  influential and universal lessons to come out of the fuzzball program is that information can only be recovered from black holes if there are $\cO(1)$ corrections to the Hawking states at the horizon.  Part of the purpose of this paper is to suggest other things that should be universal lessons.  We have argued that there are at least two other scales that are of crucial importance in the problem:  The transition length and the gap energy.  In particular, these scales limit the maximal redshift that can be observed from a microstate geometry and thus limit the temperature of the thermal fluctuations of that geometry.  Moreover, once one accepts that there must be new physical structures at the scale of the horizon then it is incumbent upon one to find a mechanism to support this structure.  Thus far, it is only the work on microstate geometries that has achieved this goal, mainly with BPS and near-BPS solutions. The mechanism behind these solutions has three incarnations, depending on the transition scale: as bubbling geometries with non-trivial fluxes, as branes polarized into other branes, and as non-Abelian effects on the world-volume of certain branes.

Finding microstate solutions for Schwarzschild-like black holes is evidently going to be extremely difficult, and at the end of the day they may not look like any of the BPS and near-BPS microstates constructed so far. Nevertheless, the arguments of \cite{Gibbons:2013tqa} and the metastable supertube physics of \cite{Bena:2011fc} make us believe that  if one wants to find {\it any} mechanism, based upon actual computations rather than insubstantial scenarios, that will support structure at the horizon of macroscopic black holes and would stand a chance of resolving the information paradox, one will be led inevitably to the mechanism described in this review. 

The black-hole microstate problem has always been one of the defining challenges for string theory.  We find it remarkable and exciting that attempts to  resolve this problem are drawing on some of the most of the innovative ideas in diverse areas of stringy research, ranging from holographic field theory, supergravity, information theory, index theory to stringy cosmology and flux stabilization.   Some more innovations will almost certainly be needed before this problem is solved but we believe that those essential innovations may soon be within reach.

\bigskip
\leftline{\bf Acknowledgements}
\smallskip
We would like to thank our collaborators over the last few years, without whom much of the progress described here would not have been possible.  IB would like to thank the Santa Barbara KITP for hospitality in the beginning stages of this work.  NPW is grateful to the IPhT, CEA-Saclay and to the Institut des Hautes Etudes Scientifiques (IHES), Bures-sur-Yvette, for hospitality while this work was completed. NPW would also like to thank the Simons Foundation for their support through a Simons Fellowship in Theoretical Physics. This work was funded in part by a grant from the Foundational Questions Institute (FQXi) Fund, a donor advised fund of the Silicon Valley Community Foundation on the basis of proposal FQXi-RFP3-1321 to the Foundational Questions Institute. This grant was administered by Theiss Research. This work was also supported in part by the ERC Starting Grant 240210 - String-QCD-BH, and by the DOE grant DE-FG03-84ER-40168, an by a grant from the FQXI.



\end{document}